\documentclass[english,aps,prb,twocolumn,showpacs,showkeys,superscriptaddress,secnumarabic,nofootinbib,floatfix]{revtex4}

\usepackage[english]{babel}

\usepackage{amssymb,amsmath} 

\usepackage[pdftex]{graphicx}
\usepackage[caption=false]{subfig}
\usepackage{color}
\usepackage[usenames,dvipsnames]{xcolor}
\usepackage[colorlinks=true,linkcolor=RoyalBlue,citecolor=OliveGreen,urlcolor=OliveGreen,linktoc=page]{hyperref} 

\usepackage{enumerate}
\usepackage{grffile}
\usepackage{balance}
\usepackage{float}

\newcommand{\Tc}{T_{\rm\scriptscriptstyle C}}
\newcommand{\Jc}{J_{\rm\scriptscriptstyle C}}
\newcommand{\Ec}{E_{\rm\scriptscriptstyle C}}
\newcommand{\Hct}{H_{\rm{\scriptscriptstyle C}2}}
\newcommand{\Jdp}{J_{\rm dp}} 
\newcommand{\Jext}{J}

\begin{document}

\title{Simulation of the vortex dynamics in a real pinning landscape \\ of YBa$_2$Cu$_3$O$_{7-\delta}$ coated conductors}

\author{I.\,A.\,Sadovskyy} 
\affiliation{Materials Science Division, Argonne National Laboratory, 9700 S. Cass Av., Argonne, IL 60637, USA}

\author{A.\,E.\,Koshelev}
\affiliation{Materials Science Division, Argonne National Laboratory, 9700 S. Cass Av., Argonne, IL 60637, USA}

\author{A.\,Glatz}
\affiliation{Materials Science Division, Argonne National Laboratory, 9700 S. Cass Av., Argonne, IL 60637, USA}
\affiliation{Department of Physics, Northern Illinois University, DeKalb, IL 60115, USA}

\author{V.\,Ortalan}
\affiliation{School of Materials Engineering, Purdue University, West Lafayette, IN 47907, USA}

\author{M.\,W.\,Rupich}
\affiliation{American Superconductor Corporation, Westborough, MA 01581, USA}

\author{M.\,Leroux}
\affiliation{Materials Science Division, Argonne National Laboratory, 9700 S. Cass Av., Argonne, IL 60637, USA}

\date{\today}

\begin{abstract}
The ability of high-temperature superconductors (HTSs) to carry very large currents with almost no dissipation makes them irreplaceable for high-power applications. The development and further improvement of HTS-based cables requires an in-depth understanding of the superconducting vortex dynamics in presence of complex pinning landscapes. We present a critical current analysis of a {\it real} HTS sample in a magnetic field by combining state-of-the-art large-scale Ginzburg-Landau simulations with reconstructive three-dimensional scanning transmission electron microscopy tomography of the pinning landscape in Dy-doped YBa$_2$Cu$_3$O$_{7-\delta}$. This methodology provides a unique look at the vortex dynamics in the presence of a complex pinning landscape, responsible for the high current-carrying capacity characteristic of commercial HTS wires. Our method demonstrates very good functional and quantitative agreement of the critical current between simulation and experiment, providing a new predictive tool for HTS wires design.
\end{abstract}



\maketitle

\section{Introduction}

Commercial high-temperature superconducting (HTS) wires are being successfully applied in a variety of electric power equipment for the power grid. Their high current-carrying capacity and low dissipation provide multiple advantages over conventional conductors.\cite{Malozemoff:2012,Shiohara:2012} However, even commercial second-generation HTS wires have large headroom for improvement.\cite{Senatore:2014,Obradors:2014} In particular, the critical current in these YBa$_2$Cu$_3$O$_{7-\delta}$ (YBCO) HTS wires decreases rapidly in magnetic fields, prohibiting them from use in broader applications such as in superconducting electrical power generators that could enable lightweight and efficient compact systems, e.g., for wind turbines. In this work we present the first vortex pinning simulation in a reconstructed mixed-pinning landscape obtained by three-dimensional (3D) STEM tomography of an actual section of a Dy-doped YBCO wire, enabling a new strategy for optimizing the critical current in HTS wires.

Energy dissipation in superconductors in the presence of an applied field arises from the motion of vortices driven by the current-induced Lorentz force,\cite{Bardeen:1965} thus restricting their mobility through pinning by admixed inclusions is the main route to minimize dissipation and increase the critical current.\cite{Foltyn:2007,Holesinger:2008,Maiorov:2009,Jia:2013} At present, the quest for higher critical current in HTS is carried out mostly via the laborious process of empirical trial and error. (Only few systematic studies of the critical current dependence on sizes and densities of defects have been published.\cite{Miura:2013a,Selvamanickam:2015,Sadovskyy:2015b}) However, at the fundamental level, the basic principles of vortex pinning have been established, at least for simple idealized situations, see, e.g., Refs.~\onlinecite{Blatter:1994,Brandt:1995,Blatter:2003}. A major impediment to rapid progress in improving the performance of superconducting wires for applications is an insufficient understanding of vortex dynamics in the complex pinning landscape of real materials.

Vortex pinning is a complicated collective phenomenon controlled by the interaction of vortices with pinning centers as well as the flexibility of vortex lines and inter-vortex interactions. The analytical treatment of this problem has been limited to qualitative estimates of the critical current in simple defect environments such as weak pinning by large densities of atomic impurities\cite{Larkin:1979} or strong pinning by low densities of strong inclusions.\cite{Ovchinnikov:1991,Blatter:2004} On the other hand, numerical simulations of vortex dynamics have been used to provide a better insight into the pinning mechanisms and to improve the quantitative characterization. Here, the choice among several models is determined by a trade-off between complexity and fidelity. For example, the minimal approach is to consider only the vortex degrees of freedom and treat vortices as particles (in thin films) or elastic strings (in the bulk). In this case, the dynamics is described by an overdamped equation of motion, which takes into account thermal Langevin forces, see, e.g., Refs.~\onlinecite{Ertas:1996,Bustingorry:2007,Luo:2007,Koshelev:2011}. This approach provides a reasonable description of vortex dynamics at small magnetic fields and small densities of pinning sites. However, vortex-vortex and vortex-defect interactions are treated only approximately in the Langevin dynamics approach and vortex cuttings and reconnections cannot be described within this model. 

\begin{figure}
	\begin{center}
		\subfloat{\includegraphics[width=8.3cm]{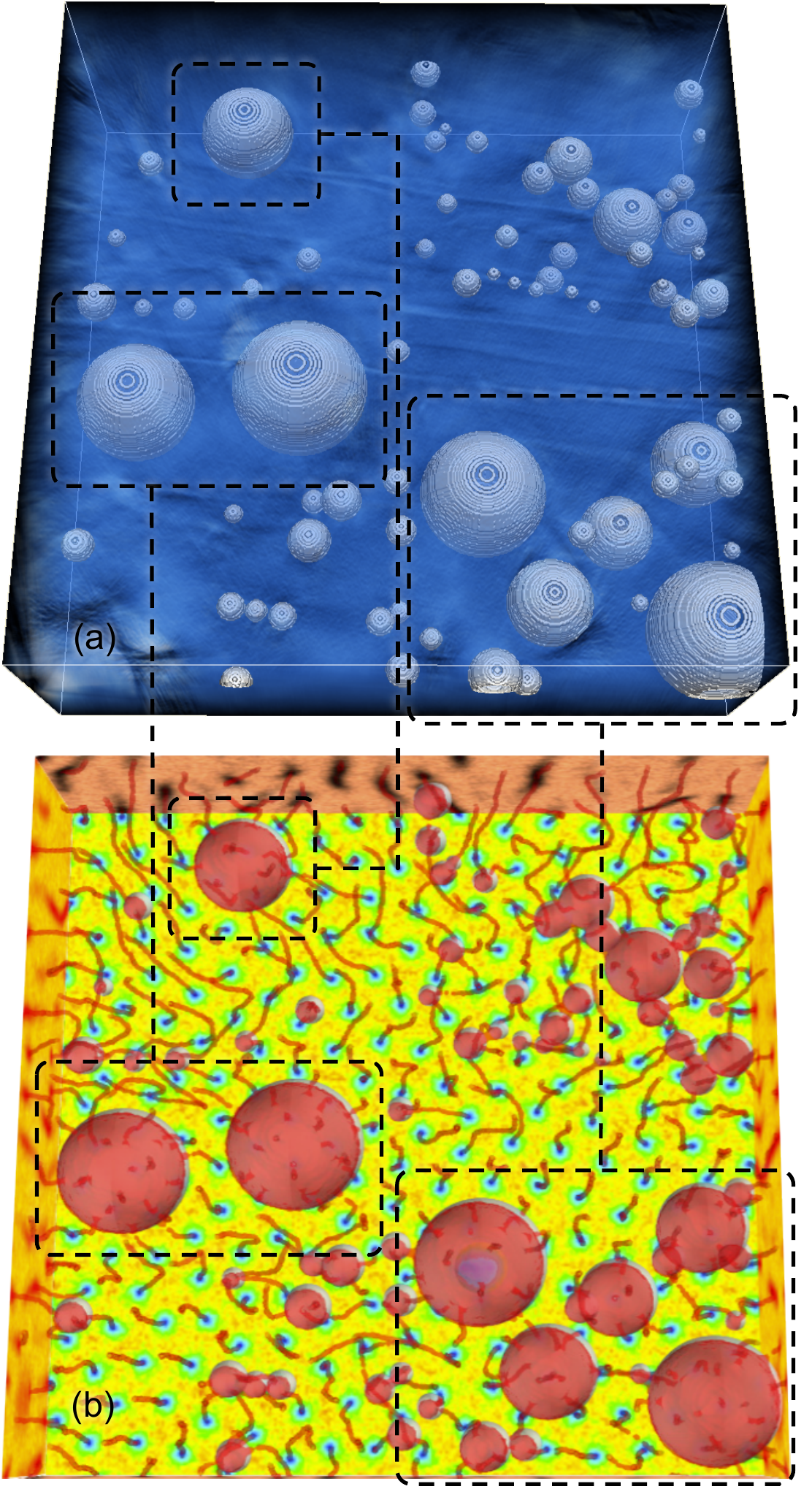} \label{fig:setup}}
		\subfloat{\label{fig:vortices}}
	\end{center} \vspace{-5mm}
	\caption{ 
		(a)~3D STEM tomogram of a Dy-doped YBCO sample 
		(doping level 0.5) processed by the non-uniform illumination 
		method superimposed with the reconstructed nanoparticles 
		by the IMOD software.\cite{Ortalan:2009} 
		(b)~Snapshot of the TDGL vortex configuration at an applied 
		magnetic field $B = 0.1\Hct = 2$\,T and external current 
		$\Jext = 0.0052\Jdp = 0.2$\,MA/cm$^2$. 
		Isosurfaces of the order parameter close to the normal 
		state are shown in red and follow both vortex and defect 
		positions. The amplitude of the order parameter on the back 
		plane of the volume is represented with a blue (normal) to 
		yellow (superconducting) color.
	}
\end{figure}

In contrast, the time-dependent Ginzburg-Landau (TDGL) approach\cite{Schmid:1966,Aranson:2002} provides an ideal compromise between approximate phenomenological and exact microscopic descriptions of vortex matter. The TDGL equations automatically account for the flexibility of the vortex lines, the long-range mutual vortex repulsion, vortex cutting and reconnection. Pinning defects of arbitrary shape and size can be incorporated via spatial modulation of the critical temperature. Furthermore, the interruption of current paths by replacement of a current-carrying superconductor with non-superconducting pinning defects is automatically accounted for by the model. TDGL-based numerical simulations have been used several times in the past to study various properties of vortex matter.\cite{Doria:1990,Machida:1993,Crabtree:1996,Aranson:1996,Crabtree:2000,Winiecki:2002,Vodolazov:2013,Berdiyorov:2014} 

Recent advancements in computational capabilities in combination with efficient parallel solvers for the TDGL equation on modern graphics processing units (GPUs)\cite{Sadovskyy:2015a} enabled the simulation of rather large 3D samples with different pinning landscapes making meaningful predictions for the behavior of critical currents possible. Practically all theoretical and numerical studies of vortex pinning dealt with idealized models where only one type of pinning center is typically considered. However, commercial HTS wires have been engineered with a complex variety of pinning defects of different sizes and shapes, which have empirically been found beneficial for high critical current densities. A direct simulation of the vortex behavior in such mixed landscape is usually not possible because information about internal structures remains qualitative and scarce. A recent scanning transmission electron microscopy (STEM) tomography study,\cite{Ortalan:2009} has determined the location and size of inclusions within a superconducting Dy-doped YBCO compound used for second-generation HTS wires, see Fig.~\subref{fig:setup}. This technique allows for unprecedented exact mapping of inclusion sizes and locations within the sample.

Here, we combine the advances in large-scale TDGL simulations and STEM tomography information to produce the first comparison of the computed critical currents for a realistic sample with experimental measurements at different magnitudes and orientations of the external magnetic field. We simulate the complete volume and nanoparticles pinning structure of the experimental system down to the resolution limit of the tomography.

\section{Model}

We use TDGL equations as the main tool for the numerical analysis of vortex dynamics. In the infinite-$\lambda$ limit, these equations provide a quantitatively adequate description of strong {type-II} superconductors at high magnetic fields. In this limit, the Maxwell-type equation for the vector potential is eliminated and we concentrate on the remaining equation for the superconducting order parameter $\psi = \psi({\bf r},t)$,
\begin{equation} 
	(\partial_t + i\mu)\psi 
	= \epsilon ({\bf r})\psi - |\psi |^2\psi + 
	(\nabla - i{\bf A})^2\psi + \zeta({\bf r},t),
	\label{eq:GL} 
\end{equation} 
where $\mu = \mu({\bf r},t)$ is the chemical potential, $\bf A$ is the vector potential associated with the external magnetic field ${\bf B}$ as ${\bf B} = \nabla \times {\bf A}$, and $\zeta({\bf r},t)$ is the temperature-dependent $\delta$-correlated Langevin term.\cite{Sadovskyy:2015a} In Eq.~\eqref{eq:GL}, written in the dimensionless form, the unit of length is given by the superconducting coherence length~$\xi$, the unit of time is $t_0 \equiv 4\pi \sigma\lambda^2/c^2$, where $\lambda$ is the London penetration depth, $\sigma$ the normal state conductance, and the unit of magnetic field is given by the $c$-axis upper critical field $\Hct = \hbar c / 2e \xi^2$. Here $-e$ is the electron's charge and $c$ the speed of light. In our notations of the TDGL equations, the $ab$-plane of the HTS is in the $xy$-plane of the simulation and the $c$-axis is along the $z$-direction.

To account for the anisotropy and layered structure of the material, we replace the $z$-component of the Laplacian term in Eq.~\eqref{eq:GL}, $(\nabla_z - i A_z)^2 \psi$, with the discrete second derivative $[\psi(z + h_z)e^{-i A_z h_z} + \psi(z - h_z)e^{i A_z h_z} - 2\psi(z)] / \gamma^2 h_z^2$, where $\gamma$ is the anisotropy factor and $h_z$ is the grid point spacing in $z$-direction, $h_z \equiv L_z / N_z$. While similar numerical discretizations are used for the in-plane directions ($x$ and $y$), the discreteness in $z$ direction is an essential part of the model. This description is known as the Lawrence-Doniach model.\cite{Lawrence:1970} In this model the effective interlayer coupling is controlled by the parameter $\xi / \gamma h_z$. In particular, the layered structure leads to intrinsic pinning of vortices for magnetic fields applied in-plane.

The dimensionless function $\epsilon({\bf r}) \propto \Tc( {\bf r}) - T$ vanishes at the local critical temperature $T \to \Tc({\bf r})$. The explicit dependence of the critical temperature~$\Tc({\bf r})$ on spatial coordinate~${\bf r}$ is used to model large-scale inhomogeneities in the superconductor and is a convenient way to introduce pinning effects. We use $\epsilon({\bf r}) = 1$ inside the superconductor. This corresponds to setting the coherence length $\xi$ at a given temperature as the unit of length. Inside non-superconducting inclusions we choose $\epsilon({\bf r}) = -1$.

The total (normal and superconducting) in-plane reduced current density in units of $J_0 \equiv \hbar c^2 / 8\pi e \lambda^2 \xi$ is given by the expression 
\begin{equation}
	{\bf J} 
	= {\rm Im} \bigl[ \psi^*(\nabla - i{\bf A})\psi \bigr]
	- \nabla \mu.
	\label{eq:J} 
\end{equation} 
The maximum theoretical depairing current density is then $\Jdp = 2/(3\sqrt{3}) J_0$. In addition to Eq.~\eqref{eq:GL}, we solve the Poisson equation, $\nabla {\bf J}=0$, for the scalar potential~$\mu$. 

To determine the critical current value, $\Jc$, we apply an external current, $\Jext$, in the $ab$-plane (in the $x$-direction) and ramp it down from the resistive to the superconducting state of the sample. Each time step we calculate the electric field in direction of the applied current averaged across the sample, i.e., $E = \langle\nabla_x \mu\rangle$. The $I$-$V$ curve is then obtained by averaging the electric field $E$ over the steady states reached after the transient regime following each current ramping event. We use a finite electric field criterion $\Ec = 10^{-4} E_0$ to determine the critical current, where $E_0 \equiv J_0/\sigma$ is the electric field unit. In other words, we compare electrical field $E$ induced by the external current $\Jext$ with a certain electric field level $\Ec$. The defined level is sufficiently low for a reasonable definition of the critical current. However, it is much higher than the level of dissipation corresponding to the value of $1\,\mu$V/cm, routinely used as a practical criterion for $\Jc$. Therefore, the simulated critical currents are expected to be somewhat higher than the experimental ones. 

To simulate the anisotropy of the critical current we apply and rotate the magnetic field ${\bf B}$ from the $ab$-plane to the $c$-axis of the HTS keeping it perpendicular to the external current. In the coordinate system of the simulation, the applied field ${\bf B} = B \bigl[0, \; \sin\theta, \; \cos\theta \bigr]$ can be described by the gauge ${\bf A} = x B \bigl[0, \; \cos\theta, \; - \sin\theta \bigr]$, where~$B$ is the absolute value of the field and $\theta$ the angle with respect to the $c$-axis. 

In the simulated volume, pinning centers are positioned in strict accordance with their positions and sizes as measured by 3D STEM tomography\cite{Ortalan:2009} in the experimental sample. The original reconstruction of the superconductor volume together with the particles obtained by the segmentation of the STEM tomogram using the IMOD software is shown in Fig.~\subref{fig:setup}. The defect inclusions are confined within a rectangular box of size $534 \times 524 \times 129$\,nm$^3$. This box contains 71 almost spherical particles with sizes ranging from 12.2 to 100\,nm. For the numerical analysis we use a coherence length $\xi = 4.2$\,nm in the $ab$-plane, which is close to the experimental value at 77\,K and an anisotropy factor of $\gamma = 5$ suitable for YBCO. We simulate a volume of size $L_x \times L_y \times L_z = 128\xi \times 128\xi \times 32\xi$, which corresponds to $538 \times 538 \times 134$\,nm$^3$ with our choice of~$\xi$. The non-superconducting particles with diameters ranging from 2.90 to $23.8\xi$ occupy about 8.1\% of the simulated volume. A typical vortex configuration for magnetic field $B = 0.1\Hct = 2$\,T is shown in Fig.~\subref{fig:vortices}, where $\Hct = 20$\,T is the $c$-axis upper critical field at 77\,K. Isosurfaces $|\psi|^2 = 0.1$ of the order parameter, shown in red in Fig.~\subref{fig:vortices}, reveal the vortex positions and the contours of the pinning landscape. On the back and sides of the simulated volume, a color-code indicates the amplitude of $|\psi|^2$.

\section{Results}

In Fig.~\ref{fig:Jc_B}, we present dependence of the critical current, $\Jc(B)$, on the magnetic field applied parallel to the $c$-axis of the HTS ($\theta = 0^\circ$). Experimentally,\cite{Ortalan:2009} at $77$\,K and in a magnetic field ranging from 0.02 to $1.5$\,T the critical current follows a power-law dependence $\Jc \propto B^{-\alpha}$ with an exponent decreasing with increasing Dy doping $\alpha \approx 0.74$ (green curve) and $0.69$ (yellow curve) for Dy doping $0.5$ and $0.75$ respectively.\footnote{Note, that we refitted the experimental data in the same field range as for the simulation data. Therefore the exponents are slightly different from the ones reported in Ref.~\onlinecite{Long:2005}.} At lower magnetic field, $B \lesssim 0.01$\,T, the critical current is more or less independent of the external field, since self-field effects dominate. In this work we concentrate on higher-field region, $B \gtrsim 0.005\Hct = 0.01$\,T, where this effect is negligible. The red curve in Fig.~\ref{fig:Jc_B} presents the simulated dependence $\Jc(B)$ from the 0.5 Dy doped sample shown in Fig.~\subref{fig:setup}, using only the STEM-resolved nanoparticles. For $B$ ranging from $0.005\Hct$ (0.1\,T) to $0.075\Hct$ (1.5\,T), corresponding to the range used in the experimental measurements, the simulated $\Jc(B)$ is also described by a power law but with the slightly larger exponent, $\alpha \approx 0.80$. To show the effects of defects smaller than the resolution of the tomography, we also added 1500 spherical particles of size $2\xi$ (8.4\,nm) to the reconstructed defects. These background inclusions reduce the exponent of the $\Jc(B)$ dependence to $\alpha \approx 0.68$, see blue curve in Fig.~\ref{fig:Jc_B}. 

In Fig.~\ref{fig:Jc_angle}, we present the anisotropy of the critical current in a tilted magnetic field, always perpendicular to the applied current. The dependencies $\Jc(\theta)$ on the tilt angle of the magnetic field $\theta$ with respect to the $c$-axis for three different magnetic fields $B$ are shown. The red curves correspond to the simulation with only the defects detected by STEM tomography, while the blue lines correspond to the simulation with the additional background inclusions. The anisotropy of $\Jc(\theta)$ increases with magnetic field and at the highest fields, $B \gtrsim 0.01\Hct$, $\Jc$ is nearly flat at all angles except for a narrow peak around $\theta = 90^\circ$ caused by the simulated intrinsic pinning due to the layered structure of the material. The ratio $\Jc(90^\circ)/\Jc(0^\circ)$ varies depending on the field: $\Jc(90^\circ)/\Jc(0^\circ) \approx 2.9$ for $B=0.005\Hct$ [Fig.~\subref{fig:Jc_angle_B=0p005}], $\approx 3.9$ for $B=0.05\Hct$ [Fig.~\subref{fig:Jc_angle_B=0p05}], and $\approx 5.3$ for $B=0.1\Hct$ [Fig.~\subref{fig:Jc_angle_B=0p1}]. This ratio is mainly determined by the Lawrence-Doniach parameter $\xi/\gamma h_z = 0.4$.

\section{Discussions}

\begin{figure}[tb]
	\begin{center} \hspace{-1.3mm}
		\includegraphics[width=8.8cm]{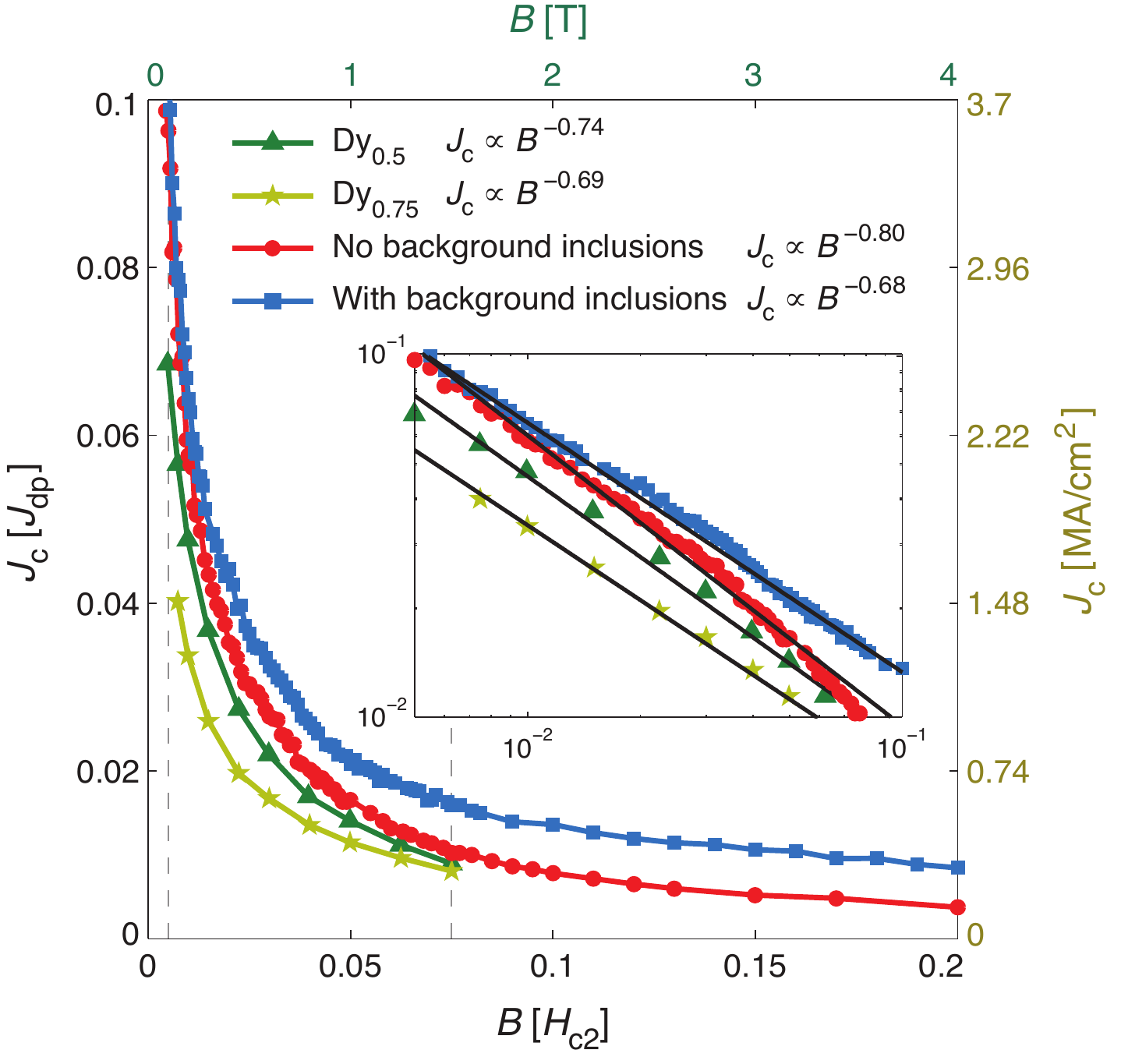}
	\end{center} \vspace{-5mm} 
	\caption{
		Critical current $\Jc$ as a function of the magnetic field $B$ applied 
		along the $c$-axis of YBCO. The simulated field dependence 
		(red curve) with only the nanoparticles observed by STEM 
		tomography in the sample with 0.5 Dy doping exhibits almost 
		the same exponent as the experiment (green curve). 
		Adding small inclusions, $2\xi$ in diameter, to the simulation 
		makes the dependence less steep (blue curve), which yields 
		an exponent very similar to the experimental one in the sample 
		with 0.75 Dy doping (yellow curve).
		The experimental $\Jc(B)$ dependencies were taken from 
		Refs.~\onlinecite{Long:2005,Ortalan:2009} for~$B$ ranging 
		from 0.1\,T ($0.005 \Hct$) to 1.5\,T ($0.075 \Hct$).
		Inset: Double-logarithmic scale to show the exponents. 
	}
	\label{fig:Jc_B}
\end{figure}

The almost quantitative agreement between the $\Jc(B)$ exponents in the simulation (0.80) and the measurements (0.74) is quite striking for this first of its kind numerical simulation of a real 3D sample. The small discrepancy in the exponents is most likely related to the presence of background inclusions that are smaller than the resolution of the STEM tomogram. In fact, we have shown that those additional small defects reduce the field exponent in the simulations. Such small inclusions, that several groups are trying to control through chemical processes,\cite{Gutierrez:2007,Miura:2013a,Miura:2013b} become especially relevant at high fields when all strong pinning sites are already occupied. In this regime it was also found in experiments that their effect is to reduce the exponent of the field dependence of the critical current.\cite{Jia:2013,Leroux:2015} Specifically, the addition of $1500$ spherical nanoparticles with diameter $2\xi$, occupying only 1.2\% of the sample volume, reduces the exponent from $\alpha \approx 0.80$ (red curve in Fig.~\ref{fig:Jc_B}) to $\alpha \approx 0.68$ (blue curve). The latter essentially coincides with the experimental value $\alpha \approx 0.69$ for 0.75 Dy doping. The increased Dy content is expected to create more nanoparticles and thus to reduce the exponent $\alpha$. Our simulation therefore appears to correctly reproduce this trend.

Next, we compare the absolute values of the critical current. In the experiment,\cite{Ortalan:2009} the critical current density varies from $\Jc = 2.5$\,MA/cm$^2$ at magnetic field $B \lesssim 0.1$\,T to $\Jc = 0.4$\,MA/cm$^2$ at $B = 1$\,T, for $B$ applied along the $c$-axis. From our numerical simulations for these two fields we obtain $\Jc = 0.10\Jdp = 3.8$\,MA/cm$^2$ and $\Jc = 0.021\Jdp = 0.77$\,MA/cm$^2$, respectively. Here we estimate $\Jdp = 37$\,MA/cm$^2$ using $\xi = 4.2$\,nm and $\lambda = 270$\,nm at 77\,K. The absolute values are also in a reasonable agreement (within a factor of two) with the experiment. This semi-quantitative agreement is quite remarkable considering that we included only the Dy nanoparticles as pinning centers but neglected other types of defects present in YBCO films such as twin boundaries, atomic point defects, stacking faults, and dislocations. In addition, at least some of the larger simulated values of~$\Jc$ may be attributed to the higher critical electric field criterion used for their determination, as explained in the model section. This suggests that the nanoparticle defects play the dominant role in the field dependence of the critical current in the measured commercial tape.

\begin{figure*}
	\begin{center}
		\subfloat{\includegraphics[width=18.0cm]{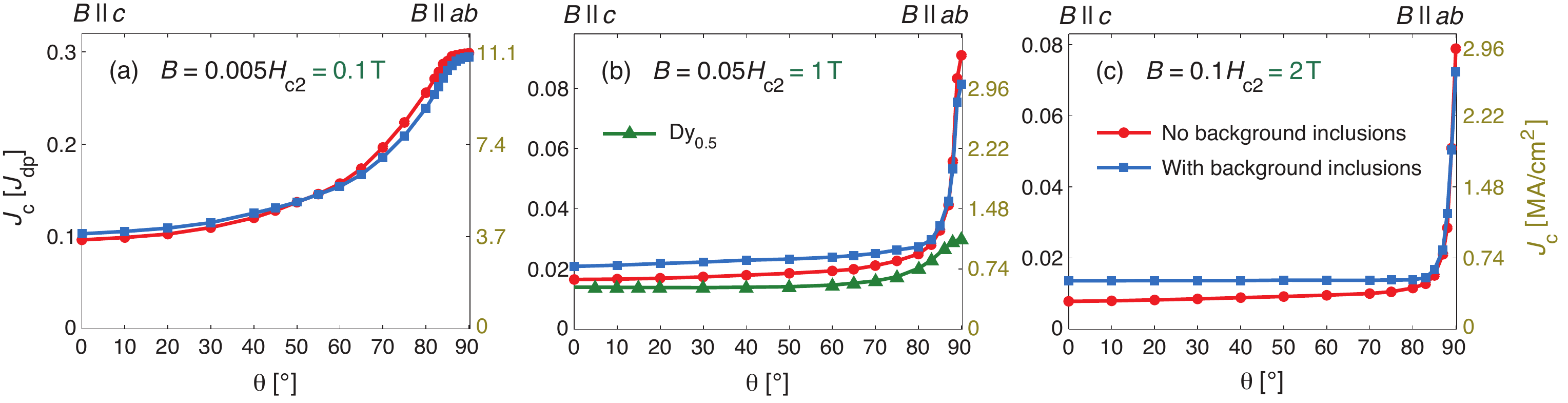} \label{fig:Jc_angle_B=0p005}}
		\subfloat{\label{fig:Jc_angle_B=0p05}}
		\subfloat{\label{fig:Jc_angle_B=0p1}}
	\end{center} \vspace{-5mm} 
	\caption{
		Simulated critical current $\Jc$ as a function of the angle $\theta$ 
		of the applied magnetic field and for different values of magnetic field: 
		(a)~$B = 0.005\Hct$, 
		(b)~$B = 0.05\Hct$ with experimental data\cite{Long:2005} 
		for 0.5 Dy doping and $B = 1$\,T superimposed, and 
		(c)~$B = 0.1\Hct$. 
	}
	\label{fig:Jc_angle}
\end{figure*}

As shown in Fig.~\subref{fig:Jc_angle_B=0p05}, the overall shape of the angular dependence appears consistent with the experiment,\cite{Ortalan:2009} in particular the flat plateau around $\theta = 0^\circ$ ($B||c$) is clearly reproduced. The overall shape of angular dependences is expected. As the nanoparticles have an isotropic shape and no $c$-axis correlated pinning centers (e.g. dislocations or twin boundaries) were introduced, there is no sharp peak in $\Jc$ for $B||c$. The only correlated pinning is the intrinsic pinning in the $ab$-plane which does yield a peak in $\Jc$. In addition, in an anisotropic material $\Jc$ is expected to have smooth angular dependence with the typical angle $\theta \sim \arctan{\gamma}$, as we observe at small field, Fig.~\subref{fig:Jc_angle_B=0p005}. With increasing magnetic field, the simulated $\Jc$ anisotropy is found to increase, a trend also in agreement with experiments.\cite{Holesinger:2008} In absolute terms, the simulated anisotropy is $\Jc(90^\circ)/\Jc(0^\circ) \approx 3.9$ for $B = 1$\,T, whereas the experimental anisotropy is somewhat smaller $\approx 2.2$ for 0.5 Dy doping and $\approx 4.5$ for the undoped sample. This discrepancy is easily explained since our choice of the Lawrence-Doniach parameter $\xi/\gamma h_z = 0.4$ yields qualitatively reasonable $\Jc$ anisotropy but would need to be fine-tuned against experimental measurements of the intrinsic pinning alone in clean single crystal for really quantitative agreement. More importantly, beyond intrinsic pinning we did not take into account other types of $ab$-plane pinning centers, such as flat precipitates, as they do not appear in the tomogram. Nonetheless, our model of $ab$-plane pinning is reasonable on a qualitative level, and the simulated angular dependence does exhibit the proper trends for the combined effects of nanoparticles and intrinsic pinning.

Finally, from a practical point of view, the most relevant question is probably how ``optimal'' the observed pinning configuration is in terms of absolute critical current values. The question can be answered in part by looking at a recent study,\cite{Koshelev:2015} by some of the present authors, on monodisperse spherical defects using the same approach. It was found that the optimal critical current for such defects, in magnetic fields ranging from $0.05\Hct$ to $0.1\Hct$, was achieved for defect diameters ranging between $3.5\xi$ to $4\xi$ and occupying a volume fraction of about $20\%$. In the sample measured by tomography, the nanoparticles occupy a total volume fraction of $8.1$\%. The distribution of diameters of these particles peaks in the range $3\xi$ to $7\xi$ (72\% of all particle are in this range) [see histogram Fig.~5(d) in Ref.~\onlinecite{Ortalan:2009}]. However, the volume fraction occupied by the above mentioned ``optimal'' particles is only $0.8\%$. Even counting all particles from $2\xi$ to $7\xi$ and taking into account the added background inclusions, barely increases the volume fraction of these pinning centers to 2\%, which is ten times less than the optimal density found in Ref.~\onlinecite{Koshelev:2015}. Most of the defect volume fraction is actually due to a few large nanoparticles, which may capture several vortices at a time [see Fig.~\subref{fig:vortices}]. According to our simulations, these large defects contribute little to pinning and also reduce the effective cross-section of the sample, thus negatively affecting the critical current.

In fact, the critical currents found in this study are about three times smaller than for the optimal pinning configuration for monodisperse spherical defects mentioned above. This therefore suggests that the present configuration is far from optimal and could be improved by skewing the distribution of nanoparticles toward smaller sizes, for instance 10--20\,nm in diameter, and simultaneously by raising their density. This is in line with recent experimental results that find extremely effective pinning with nanoparticles of size 15--30\,nm\cite{Gutierrez:2007} or 20--80\,nm\cite{Miura:2013a} in YBCO, and 8\,nm in pnictides.\cite{Miura:2013b} Also, similar results were observed with smaller irradiation-induced defects, 5\,nm in size.\cite{Jia:2013,Leroux:2015}

\section{Conclusions}

In summary, we numerically simulated the superconducting vortex dynamics in a real pinning landscape of nanoparticles inside a commercial Dy-doped YBCO tape, using the same sample size as in the experiment. The positions and sizes of these nanoparticles were obtained directly through STEM tomography reconstruction of the sample nanostructure. We obtained good qualitative and almost quantitative agreement in the functional dependencies and absolute values of the critical current between simulation and experiment. One can expect that more detailed STEM tomography studies, and in particular more quantitative characterizations of pinning centers, should improve the quantitative agreement. In addition, there is still room for improving the simulation in terms of exact treatment of intrinsic pinning as well as lowering the voltage criterion towards the one used in the industry. Our results, however, show a promising pathway for the quantitative analysis and optimization of vortex dynamics in various realistic defects environment and physical conditions, using a combination of 3D tomography analysis and large-scale time-dependent Ginzburg-Landau simulations.

\subsection*{Acknowledgements} 

We are delighted to thank A.\,P.\,Malozemoff, W.-K. Kwok, and U.\,Welp for careful reading of the manuscript and numerous useful comments. This work was supported by the Scientific Discovery through Advanced Computing (SciDAC) program funded by U.S. Department of Energy Office of Science, Advanced Scientific Computing Research. Simulations were performed at Argonne LCF supported by DOE under contract DE-AC02-06CH11357. A.\,E.\,K. and M.\,L. acknowledge support by the Center for Emergent Superconductivity, an Energy Frontier Research Center funded by the U.S. Department of Energy, Office of Science, Office of Basic Energy Sciences Award No. DEAC0298CH1088.

\bibliography{tomogram}

\end{document}